\pdfoutput=1
\documentclass[12pt,a4paper]{article}

\usepackage{ifthen} 
\newboolean{pdflatex}
\setboolean{pdflatex}{true} 

\newboolean{articletitles}
\setboolean{articletitles}{true} 

\newboolean{uprightparticles}
\setboolean{uprightparticles}{false} 

\usepackage{booktabs}

\def\paperauthors{LHCb collaboration} 
\def\paperasciititle{First observation of the decay Bs -> K- mu+ nu and a measurement of |Vub|/|Vcb|} 
\def\papertitle{First observation \\ of the decay $\Bs \to \Km\mup\nu_\mu$ \\and a measurement of $|V_{ub}|/|V_{cb}|$} 
\def\paperkeywords{{High Energy Physics}, {LHCb}} 
\def\papercopyright{\the\year\ CERN for the benefit of the LHCb collaboration} 
\def\paperlicence{CC BY 4.0 licence}
\def\paperlicenceurl{https://creativecommons.org/licenses/by/4.0/}

\usepackage[top=1in, bottom=1.25in, left=1in, right=1in]{geometry}

\usepackage{microtype}
\usepackage{lineno}  
\usepackage{xspace} 
\usepackage{caption} 

\usepackage{graphicx}  
\usepackage{color}
\usepackage{colortbl}
\graphicspath{{./figs/}} 

\usepackage{amsmath} 
\usepackage{amssymb}
\usepackage{amsfonts}
\usepackage{upgreek} 

\newcommand*\patchAmsMathEnvironmentForLineno[1]{%
\expandafter\let\csname old#1\expandafter\endcsname\csname #1\endcsname
\expandafter\let\csname oldend#1\expandafter\endcsname\csname
end#1\endcsname
 \renewenvironment{#1}%
   {\linenomath\csname old#1\endcsname}%
   {\csname oldend#1\endcsname\endlinenomath}%
}
\newcommand*\patchBothAmsMathEnvironmentsForLineno[1]{%
  \patchAmsMathEnvironmentForLineno{#1}%
  \patchAmsMathEnvironmentForLineno{#1*}%
}
\AtBeginDocument{%
\patchBothAmsMathEnvironmentsForLineno{equation}%
\patchBothAmsMathEnvironmentsForLineno{align}%
\patchBothAmsMathEnvironmentsForLineno{flalign}%
\patchBothAmsMathEnvironmentsForLineno{alignat}%
\patchBothAmsMathEnvironmentsForLineno{gather}%
\patchBothAmsMathEnvironmentsForLineno{multline}%
\patchBothAmsMathEnvironmentsForLineno{eqnarray}%
}

\usepackage{hyperxmp}

\usepackage[pdftex,
            pdfauthor={\paperauthors},
            pdftitle={\paperasciititle},
            pdfkeywords={\paperkeywords},
            pdfcopyright={Copyright (C) \papercopyright},
            pdflicenseurl={\paperlicenceurl}]{hyperref}

\usepackage[colorinlistoftodos,textsize=scriptsize]{todonotes}

\usepackage[bottom,flushmargin,hang,multiple]{footmisc}

\usepackage[all]{hypcap} 

\usepackage{xspace} 
\usepackage{upgreek}

\def\lhcb   {\mbox{LHCb}\xspace}

\def\MagUp {\mbox{\em Mag\kern -0.05em Up}\xspace}

\ifthenelse{\boolean{uprightparticles}}%
{

 \def\Pmu         {\ensuremath{\upmu}\xspace}

 \def\Ppi         {\ensuremath{\uppi}\xspace}

 \def\Ptau        {\ensuremath{\uptau}\xspace}

 \def\Ppsi        {\ensuremath{\uppsi}\xspace}

 \def\PDelta      {\ensuremath{\Delta}\xspace}                 
 \def\PXi         {\ensuremath{\Xi}\xspace}                 
 \def\PLambda     {\ensuremath{\Lambda}\xspace}                 
 \def\PSigma      {\ensuremath{\Sigma}\xspace}                 
 \def\POmega      {\ensuremath{\Omega}\xspace}                 
 \def\PUpsilon    {\ensuremath{\Upsilon}\xspace}

 \def\PB      {\ensuremath{\mathrm{B}}\xspace}                 
                  
 \def\PD      {\ensuremath{\mathrm{D}}\xspace}

 \def\PJ      {\ensuremath{\mathrm{J}}\xspace}                 
 \def\PK      {\ensuremath{\mathrm{K}}\xspace}

 \def\Pb      {\ensuremath{\mathrm{b}}\xspace}                 
 \def\Pc      {\ensuremath{\mathrm{c}}\xspace}

 \def\Pi      {\ensuremath{\mathrm{i}}\xspace}

 \def\Ps      {\ensuremath{\mathrm{s}}\xspace}

 \def\thebaroffset{0.0em}
}
{

 \def\Pmu         {\ensuremath{\mu}\xspace}

 \def\Ppi         {\ensuremath{\pi}\xspace}

 \def\Ptau        {\ensuremath{\tau}\xspace}

 \def\Ppsi        {\ensuremath{\psi}\xspace}                 
                  
 \mathchardef\PDelta="7101
 \mathchardef\PXi="7104
 \mathchardef\PLambda="7103
 \mathchardef\PSigma="7106
 \mathchardef\POmega="710A
 \mathchardef\PUpsilon="7107
                  
 \def\PB      {\ensuremath{B}\xspace}                 
                  
 \def\PD      {\ensuremath{D}\xspace}

 \def\PJ      {\ensuremath{J}\xspace}                 
 \def\PK      {\ensuremath{K}\xspace}

 \def\Pb      {\ensuremath{b}\xspace}                 
 \def\Pc      {\ensuremath{c}\xspace}

 \def\Pi      {\ensuremath{i}\xspace}

 \def\Ps      {\ensuremath{s}\xspace}

 \def\thebaroffset{0.18em}
}
\newcommand{\offsetoverline}[2][\thebaroffset]{\kern #1\overline{\kern -#1 #2}}%

\makeatletter
\ifcase \@ptsize \relax
  \newcommand{\miniscule}{\@setfontsize\miniscule{4}{5}}
\or
  \newcommand{\miniscule}{\@setfontsize\miniscule{5}{6}}
\or
  \newcommand{\miniscule}{\@setfontsize\miniscule{5}{6}}
\fi
\makeatother

\DeclareRobustCommand{\optbar}[1]{\shortstack{{\miniscule (\rule[.5ex]{1.25em}{.18mm})}
  \\ [-.7ex] $#1$}}

\def\mup        {{\ensuremath{\Pmu^+}}\xspace}
\def\mun        {{\ensuremath{\Pmu^-}}\xspace} 

\def\taup       {{\ensuremath{\Ptau^+}}\xspace}

\def\squark    {{\ensuremath{\Ps}}\xspace}

\def\cquark    {{\ensuremath{\Pc}}\xspace}

\def\bquark    {{\ensuremath{\Pb}}\xspace}

\def\pion   {{\ensuremath{\Ppi}}\xspace}
\def\piz    {{\ensuremath{\pion^0}}\xspace}

\def\pim    {{\ensuremath{\pion^-}}\xspace}

\def\kaon    {{\ensuremath{\PK}}\xspace}

\def\KorKbar {\kern \thebaroffset\optbar{\kern -\thebaroffset \PK}{}\xspace}

\def\Kp      {{\ensuremath{\kaon^+}}\xspace}
\def\Km      {{\ensuremath{\kaon^-}}\xspace}

\def\D       {{\ensuremath{\PD}}\xspace}

\def\DorDbar {\kern \thebaroffset\optbar{\kern -\thebaroffset \PD}\xspace}

\def\Dp      {{\ensuremath{\D^+}}\xspace}
\def\Dm      {{\ensuremath{\D^-}}\xspace}

\def\DpDm    {\ensuremath{\Dp {\kern -0.16em \Dm}}\xspace}

\def\Dsm     {{\ensuremath{\D^-_\squark}}\xspace}

\def\B       {{\ensuremath{\PB}}\xspace}

\def\BorBbar {\kern \thebaroffset\optbar{\kern -\thebaroffset \PB}\xspace}

\def\Bd      {{\ensuremath{\B^0}}\xspace}

\def\BdorBdbar {\kern \thebaroffset\optbar{\kern -\thebaroffset \Bd}\xspace}

\def\Bs      {{\ensuremath{\B^0_\squark}}\xspace}

\def\BsorBsbar {\kern \thebaroffset\optbar{\kern -\thebaroffset \Bs}\xspace}

\def\jpsi     {{\ensuremath{{\PJ\mskip -3mu/\mskip -2mu\Ppsi}}}\xspace}
\def\psitwos  {{\ensuremath{\Ppsi{(2S)}}}\xspace}

\def\Y#1S{\ensuremath{\PUpsilon{(#1S)}}\xspace}

\def\Lz          {{\ensuremath{\PLambda}}\xspace}

\def\LorLbar     {\kern \thebaroffset\optbar{\kern -\thebaroffset \PLambda}\xspace}

\def\Lc          {{\ensuremath{\Lz^+_\cquark}}\xspace}

\def\Lb           {{\ensuremath{\Lz^0_\bquark}}\xspace}

\def\to                 {\ensuremath{\rightarrow}\xspace}

\def\qsq       {{\ensuremath{q^2}}\xspace}

\def\AT#1     {\ensuremath{A_{\mathrm{T}}^{#1}}\xspace}           

\def\C#1      {\ensuremath{\mathcal{C}_{#1}}\xspace}                       
\def\Cp#1     {\ensuremath{\mathcal{C}_{#1}^{'}}\xspace}                    
\def\Ceff#1   {\ensuremath{\mathcal{C}_{#1}^{\mathrm{(eff)}}}\xspace}        
\def\Cpeff#1  {\ensuremath{\mathcal{C}_{#1}^{'\mathrm{(eff)}}}\xspace}       
\def\Ope#1    {\ensuremath{\mathcal{O}_{#1}}\xspace}                       
\def\Opep#1   {\ensuremath{\mathcal{O}_{#1}^{'}}\xspace}                    

\newcommand{\aunit}[1]{\ensuremath{\text{\,#1}}}       

\newcommand{\tev}{\aunit{Te\kern -0.1em V}\xspace}
\newcommand{\gev}{\aunit{Ge\kern -0.1em V}\xspace}
\newcommand{\mev}{\aunit{Me\kern -0.1em V}\xspace}
\newcommand{\kev}{\aunit{ke\kern -0.1em V}\xspace}
\newcommand{\ev}{\aunit{e\kern -0.1em V}\xspace}
 
\newcommand{\mevc}{\ensuremath{\aunit{Me\kern -0.1em V\!/}c}\xspace}
\newcommand{\gevc}{\ensuremath{\aunit{Ge\kern -0.1em V\!/}c}\xspace}
\newcommand{\mevcc}{\ensuremath{\aunit{Me\kern -0.1em V\!/}c^2}\xspace}
\newcommand{\gevcc}{\ensuremath{\aunit{Ge\kern -0.1em V\!/}c^2}\xspace}
\newcommand{\gevgevcccc}{\ensuremath{\gev^2\!/c^4}\xspace} 

\def\fb   {\ensuremath{\aunit{fb}}\xspace}
\def\invfb   {\ensuremath{\fb^{-1}}\xspace}

\def\ps   {\ensuremath{\aunit{ps}}\xspace}

\def\invps{\ensuremath{\ps^{-1}}\xspace}

\newcommand{\stat}{\aunit{(stat)}\xspace}
\newcommand{\syst}{\aunit{(syst)}\xspace}

\def\gsim{{~\raise.15em\hbox{$>$}\kern-.85em
          \lower.35em\hbox{$\sim$}~}\xspace}
\def\lsim{{~\raise.15em\hbox{$<$}\kern-.85em
          \lower.35em\hbox{$\sim$}~}\xspace}

\def\pt         {\ensuremath{p_{\mathrm{T}}}\xspace}

\def\tell1  {TELL1\xspace}
\def\ukl1   {UKL1\xspace}

\usepackage{cite} 
\usepackage{mciteplus}

\usepackage{longtable} 

\begin{document}

\renewcommand{\thefootnote}{\fnsymbol{footnote}}
\setcounter{footnote}{1}


\begin{titlepage}
\pagenumbering{roman}

\vspace*{-1.5cm}
\centerline{\large EUROPEAN ORGANIZATION FOR NUCLEAR RESEARCH (CERN)}
\vspace*{1.5cm}
\noindent
\begin{tabular*}{\linewidth}{lc@{\extracolsep{\fill}}r@{\extracolsep{0pt}}}
\ifthenelse{\boolean{pdflatex}}
{\vspace*{-1.5cm}\mbox{\!\!\!\includegraphics[width=.14\textwidth]{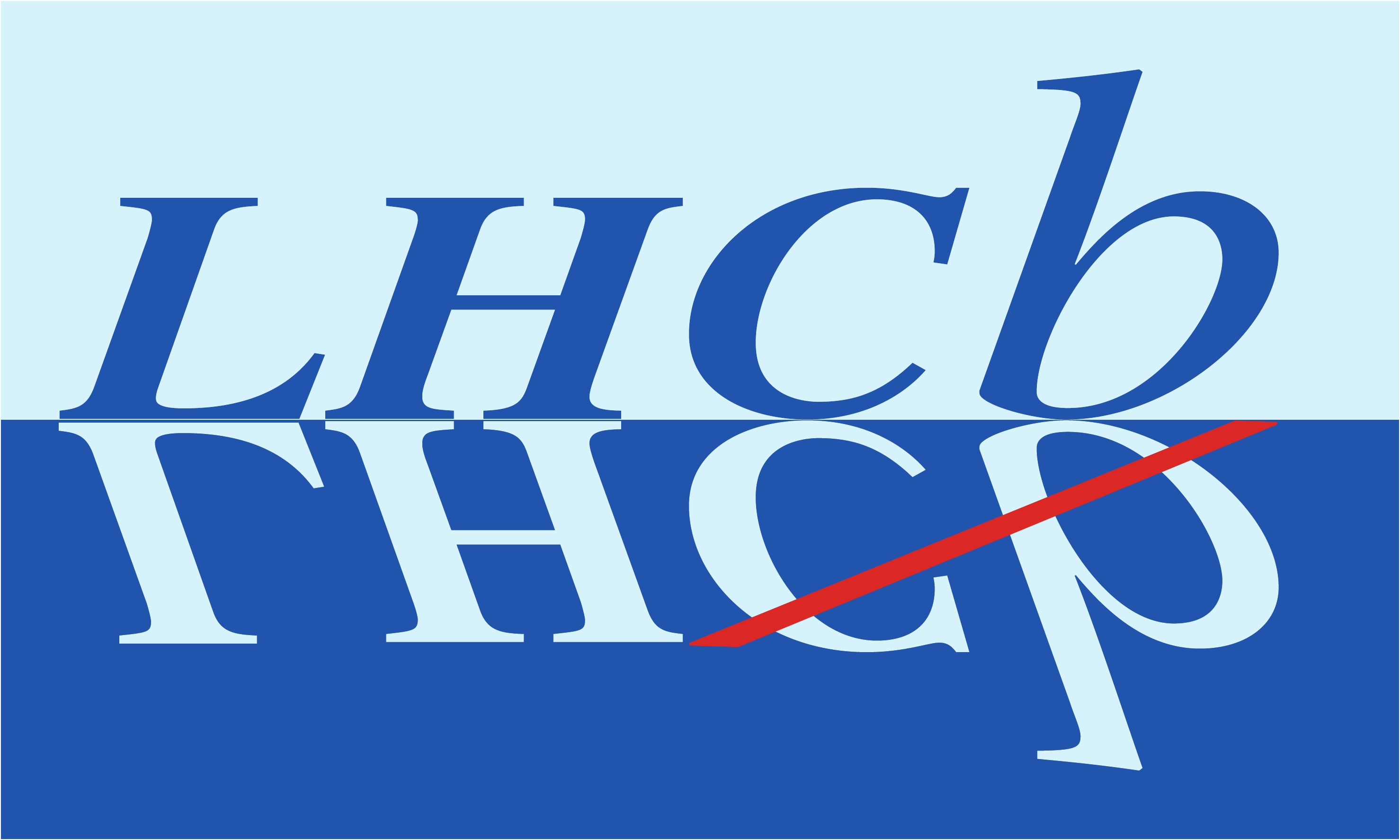}} & &}%
{\vspace*{-1.2cm}\mbox{\!\!\!\includegraphics[width=.12\textwidth]{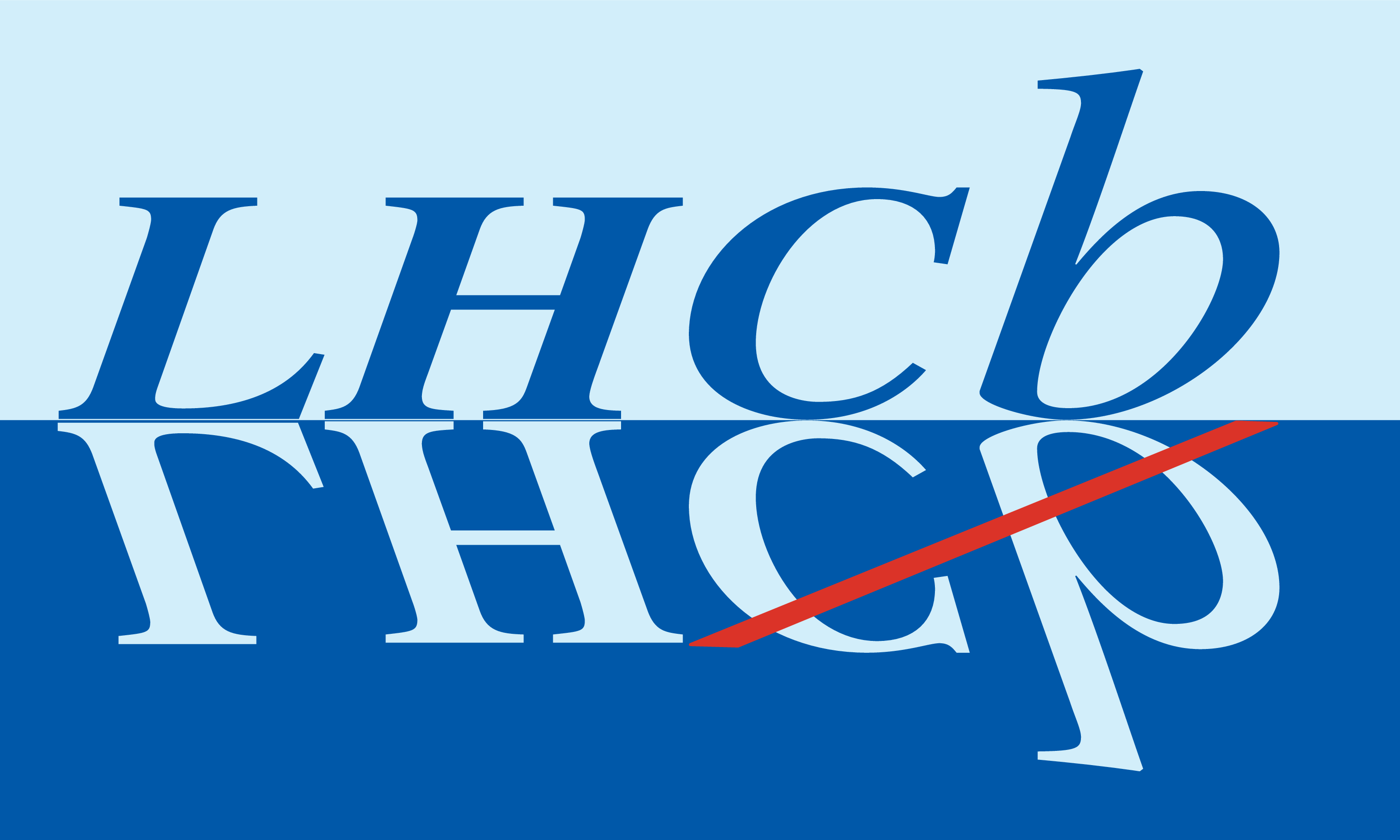}} & &}%
\\
 & & CERN-EP-2020-224 \\  
 & & LHCb-PAPER-2020-038 \\  
 & & January 15, 2021 \\ 
\end{tabular*}

\vspace*{4.0cm}

{\normalfont\bfseries\boldmath\huge
\begin{center}
  \papertitle 
\end{center}
}

\vspace*{2.0cm}

\begin{center}
\paperauthors\footnote{Authors are listed at the end of this paper.}
\end{center}

\vspace{\fill}

\begin{abstract}
  \noindent
The first observation of the suppressed semileptonic $\Bs \to \Km\mup\nu_\mu$ decay is reported. Using a data sample recorded in {\it pp} collisions in 2012 with the LHCb detector, corresponding to an integrated luminosity of 2\invfb, the branching fraction \mbox{$\mathcal{B}(\Bs \to \Km\mup\nu_\mu)$} is measured to be $(1.06\pm0.05\stat\pm0.08\syst)\times 10^{-4}$, where the first uncertainty is statistical and the second one represents the combined systematic uncertainties. The decay $\Bs \to \Dsm\mup\nu_\mu$, where $\Dsm$ is reconstructed in the final state $\Kp\Km\pim$, is used as a normalization channel to minimize the experimental systematic uncertainty. Theoretical calculations on the form factors of the $\Bs \to K^-$ and $\Bs \to \Dsm$ transitions are employed to determine the ratio of the CKM matrix elements ${|V_{ub}|}/{|V_{cb}|}$ at low and high $\Bs \to K^-$ momentum transfer.
\end{abstract}

\vspace*{2.0cm}

\begin{center}
 Published in Phys.~Rev.~Lett.
\end{center}

\vspace{\fill}

{\footnotesize 
\centerline{\copyright~\papercopyright. \href{\paperlicenceurl}{\paperlicence}.}}
\vspace*{2mm}

\end{titlepage}


\newpage
\setcounter{page}{2}
\mbox{~}
%
%
%
%


\renewcommand{\thefootnote}{\arabic{footnote}}
\setcounter{footnote}{0}

\cleardoublepage


\pagestyle{plain} 
\setcounter{page}{1}
\pagenumbering{arabic}


The coupling of the electroweak interaction between up- and down-type quarks is modulated by the Cabibbo-Kobayashi-Maskawa (CKM) matrix \cite{CKM1,CKM2}. Hadrons containing a $b$ quark can decay weakly via a virtual $W$ boson to semileptonic final states through the tree-level transitions $b\to c (W^*\to\ell\nu)$ and $b\to u (W^*\to\ell\nu)$, where $\ell\nu$ denotes a charged lepton and a neutrino. These transitions involve the CKM matrix elements $V_{cb}$ and $V_{ub}$, respectively, which obey the observed hierarchy $|V_{ub}|/|V_{cb}|\sim 0.1$, resulting in the transitions $b\to c\ell\nu$ being favored over $b\to u\ell\nu$. Semileptonic $b$-hadron decays are an excellent ground for measuring $|V_{cb}|$ and $|V_{ub}|$ since the factorization of the hadronic and leptonic parts of the amplitudes eases theoretical calculations \cite{Isgur:1988gb,Dugan:1990de}. Improving the precision on the measurements of the CKM elements can be exploited to probe possible deviations from the Standard Model of particle physics \cite{Battaglia:2003in}. Existing $|V_{ub}|$ and $|V_{cb}|$ measurements show a discrepancy between those performed with exclusive decays, where all the visible particles are reconstructed, and inclusive decays where only the lepton is reconstructed \cite{HFLAV}. The world average of the exclusive $|V_{ub}|$ results is dominated by $B^0\to\pim\ell^+\nu_\ell$ measurements. The \lhcb measurement using the baryonic decays $\Lb\to p\mun\bar{\nu}_\mu$ and $\Lb\to \Lc\mun\bar{\nu}_\mu$ \cite{LHCb-PAPER-2015-013} gives the ratio $|V_{ub}|/|V_{cb}|=0.079\pm0.006$, as updated in Ref.~\cite{HFLAV}. Besides the inclusive versus exclusive puzzle, measurements of $|V_{ub}|/|V_{cb}|$ are important to constrain the CKM unitarity triangle \cite{Charles:2004jd,Bona:2006ah}.\\
\indent This Letter reports the first observation of the decay $\Bs \to \Km\mup\nu_\mu$, the measurement of its branching fraction and of the ratio $|V_{ub}|/|V_{cb}|$ with $\Bs \to \Dsm\mup\nu_\mu$  as a normalization channel.\footnote{Throughout the paper, charge conjugate decays are implied.} The measurement of the branching fraction is performed in two regions of the $\Bs \to \Km$ momentum transfer or invariant mass squared of the muon and the neutrino, $q^2$, as well as integrated over the full \qsq range. The ratio $|V_{ub}|/|V_{cb}|$ is derived in the two $q^2$ regions using calculations of the form factors of the $\Bs \to \Km$ and $\Bs \to \Dsm$ transitions based on both light cone sum rule (LCSR) \cite{Colangelo:2000dp,Dominguez:2013ata} and lattice QCD (LQCD) \cite{Rothe} methods. The data sample consists of $pp$ collisions recorded by the \lhcb detector in 2012 at a center-of-mass energy of 8~\!TeV corresponding to $2\invfb$ of integrated luminosity. 
The LHCb detector is a single-arm forward spectrometer covering the pseudorapidity range $2 < \eta < 5$, described in detail in Refs.~\cite{LHCb-DP-2008-001,LHCb-DP-2014-002}.
The trigger~\cite{LHCb-DP-2012-004}  consists of a hardware stage, based on information from the calorimeter and muon systems, followed by a software stage, which reconstructs charged particles.
 Simulation, produced with software packages described in Refs.~\cite{Sjostrand:2007gs,*Sjostrand:2006za,Lange:2001uf,Allison:2006ve, *Agostinelli:2002hh}, is used to model the effects of the detector acceptance and the imposed selection requirements.\\
\indent In this analysis candidates for $\Bs \to \Km\mup\nu_\mu$ and $\Bs \to \Dsm\mup\nu_\mu$ decays are formed by combining a muon with a kaon or a $\Dsm$ candidate reconstructed through the decay $\Dsm\to \Kp\Km\pim$. The trigger and initial selection requirements are chosen to be similar between these two modes. Events are retained by the hardware trigger due to the presence of a high-\pt muon, where \pt is the momentum component transverse to the beam. The software trigger \cite{BBDT} selects partially reconstructed $B$ decays by combining a track or a $\Dsm$ candidate with a well identified muon candidate. The initial selection includes requirements on the track kinematics and quality, particle identification, as well as on the $\Bs$ candidate kinematics and decay topology. The obtained samples for each of the decays include background contributions dominated by $b$-hadron decays with additional tracks or neutral particles in the final state. For the $\Km\mup$ combinations, the main background originates from $H_b\to\mup H_c(\to K^- X)X^\prime$, where $H_{b,c}$ represents a hadron containing a $b$ or a $c$ quark and $X^{(\prime)}$ denotes unreconstructed particles. Decays to excited $K^*$ resonances, $\Bs\to K^{*-}(\to \Km\piz)\mup\nu_\mu$, and charmonium modes $B\to [c\bar{c}](\to\mup\mun)\Km X$, where $[c\bar{c}]=\jpsi,\psitwos$, are secondary background contributions. Other sources arise from $b$-hadron decays where a track is misidentified as a kaon or a muon, and random combinations of a muon and a kaon. In the $\Dsm\mup$ combinations, the main (and irreducible) source of background arises from $\Bs \to D_s^{*-}(\to \Dsm \gamma)\mup\nu_\mu$ decays. Additional contributions include decays to higher excitations of the $\Dsm$ meson, $\Bs \to D_s^{**-}(\to \Dsm X)\mup\nu_\mu$, double-charm decays of the type $B_{u,d,s}\to D_s D X$ and semitauonic $\Bs\to \Dsm\taup\nu_\tau$ decays.\\
\indent To suppress background, the $\Km\mup$ and $\Dsm\mup$ candidates are required to be isolated from other tracks in the event. A multivariate algorithm (MVA) is trained to determine if a given track originates from the candidate, or from the rest of the event (ROE). A threshold on the value of the MVA output is applied to the ROE track that appears to be the closest to the signal. For $\Km\mup$ candidates, two  boosted decision tree (BDT) classifiers \cite{Breiman,AdaBoost} are used sequentially to further reduce the remaining background. A {\it charged} BDT classifier is trained against a mixture of the main background components using, in addition to the isolation MVA output, invariant masses formed by the least isolated ROE track with respect to each of the muon or the kaon, and variables related to the $\Bs$, $\Km$ and $\mup$ kinematics. The background passing the {\it charged} BDT requirement comprises decays without an additional track, mainly of the type $H_b\to\mup H_c(\to \Km P)$, where $P$ is either a long-lived or a neutral particle. A second BDT classifier, denoted {\it neutral} BDT, involves kinematic variables of the $\Km$ and $\Bs$ candidates, the $\Bs$ vertex position and quality, the invariant mass formed by the signal kaon and any $\pi^0$ meson in its vicinity; it also exploits the asymmetry between the kaon momentum and an average momentum direction formed by neutral particles in the vicinity of the kaon. The shapes of the BDT outputs are calibrated with the decay $B^-\to \jpsi(\to\mup\mun) K^-$, which is reconstructed both as a $\Km\mup$ candidate and fully reconstructed where the least isolated track near the $\Km\mup$ pair is identified as $\mun$. Kinematic weighting accounts for data-simulation discrepancies for the training of the classifiers.\\
\indent The $\Bs$ mass is represented by the corrected mass \cite{MCorr_SLD}, defined as
\begin{equation}
	m_{\rm corr} = \sqrt{m^2_{Y\mu} + p_\perp^2/c^2} + p_\perp/c,
	\label{eq:MCORR}
\end{equation}
where $m_{Y\mu}$ is the invariant mass of the $Y\mu$ pair, with $Y=\Km$ or $\Dsm$, and $p_\perp$ is the momentum of this pair transverse to the $\Bs$ flight direction. The flight direction is defined as the vector between the positions of the primary $pp$ collision vertex and the $\Bs$ decay vertex. In order to improve the separation between the $\Bs \to \Km\mup\nu_\mu$ signal and background, the uncertainty on $m_{\rm corr}$ is required to  be $\sigma(m_{\rm corr})<100\mevcc$. The shape of $\sigma(m_{\rm corr})$ is calibrated following a similar procedure as for the BDT classifiers. To derive \qsq, the neutrino momentum is estimated using the $\Bs$ flight direction and the known $\Bs$ mass. A two-fold ambiguity resulting from this estimate is resolved by choosing the solution that is most consistent with the $\Bs$ momentum predicted by a linear regression method \cite{Ciezarek:2016lqu}. The fit to the $m_{\rm corr}$ distribution, used for the extraction of the $\Bs \to \Km\mup\nu_\mu$ signal, is performed in two \qsq regions, respectively above and below $7 \gevgevcccc$ (``high'' and ``low''), which are chosen to contain approximately the same expected signal yields.

For the $\Bs\to \Dsm\mup\nu_\mu$ decay, a fit to the invariant mass of the $\Dsm\to \Kp\Km\pim$ candidates is performed in 40 intervals of $m_{\rm corr}$ from 3000 to 6500\mevcc. This provides the $\D_s$ yield as a function of $m_{\rm corr}$ and thus subtracts the background originating from combinations of random kaon and pion tracks. The obtained $m_{\rm corr}$ distribution is fit to extract the $\Bs\to \Dsm\mup\nu_\mu$ signal yield. For the $\Bs \to \Km\mup\nu_\mu$ decay, the combinatorial background is largely reduced by applying a topological criterion: the opening angle between the directions of the $\Km$ and $\mup$ candidates in the plane transverse to the \mbox{$pp$ collision} axis is required to be less than 90 degrees. The efficiency of this requirement on the signal is 93\%, while it removes approximately 90\% of the combinatorial background.

The efficiencies of the signal and normalization channels are derived from simulation and take into account the effects of the triggers, reconstruction, selection, particle identification, isolation procedure, MVA requirements and detector acceptance. Data-driven corrections are applied to account for any mismodelling related to the kinematics, number of tracks in the event and particle identification variables. The efficiency ratio between the signal and normalization decays is $\mbox{$\epsilon_K/\epsilon_{D_s}=1.109\pm0.018$},~0.553\pm0.009$ and $0.733\pm0.009$ for $\qsq < 7\gevgevcccc$, $\qsq > 7\gevgevcccc$ and the full \qsq range, respectively. The uncertainties reflect the limited size of the simulated samples.

The fit template for the $m_{\rm corr}$ distribution of the $\Bs \to \Km\mup\nu_\mu$ signal is obtained from simulation, while the shapes for the background components are derived from either simulation or control samples. The statistical uncertainties originating from the finite samples used to obtain the templates are accounted for in the fits \cite{Barlow:1993dm}.
The main background $H_b\to H_c(\to \Km X)\mup X^\prime$, whose yield is free in the fit, is obtained with a simulated inclusive sample. The $\Bs\to K^{*-}(\to \Km\piz)\mup\nu_\mu$ background is modelled by simulating a mixture of three resonances ($K^{*-}(892)$, $K_0^{*-}(1430)$ and $K_2^{*-}(1430)$) with a substantial branching fraction to the $\Km\piz$ final state. Though the overall yield is free, the mixture is fixed to certain proportions which are varied up to a factor of 2.5 for systematic studies, according to available measurements of the decays $B^-\to K^{*-}\mup\mun$ and $B^-\to K^{*-}\eta/\phi$ \cite{PDG2020}. The impact of a possible $\Bs\to \Km\piz\mup\nu_\mu$ nonresonant decay has also been considered and found to be absorbed by the resonant mixture. The charmonium background is dominated by $B^-\to \jpsi(\to\mup\mun)\Km X$ decays, with the fraction of the $B^-\to \jpsi(\to\mup\mun)\Km$ channel exceeding 75\%. Its shape is determined with simulated $B^-\to J/\psi(\to\mup\mun)\Km X$ events while its yield is derived from the yield of the $B^-\to J/\psi(\to\mup\mun)\Km$ signal peak in data. To recover that peak from $\Km\mup$ combinations, the missing momentum of the $\mun$ is calculated from the $B^-$ flight direction and the known \jpsi mass.
The background originating from the misidentification (MisID) of a pion, proton or muon as a kaon; or a kaon, proton or pion as a muon is modelled using data samples of $h\mup$ ($\Km h$) candidates with an identical selection as for the main sample but where $h$ is a charged track which fails the kaon (muon) identification criteria.  These control samples are thus enriched in misidentified tracks of the different species. The different contributions to the kaon and muon MisID are unfolded using control samples of kinematically identified hadrons and muons \cite{LHCb-PUB-2016-021}. These samples are used to derive the probabilities that a particle belonging to a given species and with particular kinematic properties would pass the kaon or muon criteria. With this method both the $m_{\rm corr}$ shape and the yield of the MisID are constrained. 
The combinatorial background is modelled with a separate data sample where a kaon and a muon from different events are combined. The obtained pseudocandidates undergo the same selection as the signal candidates and are corrected to reproduce the kinematic properties of the standard candidates.\\
\indent The fit to the normalization channel $\Bs\to \Dsm\mup\nu_\mu$ employs shapes obtained from simulation. The  $\Bs\to \Dsm\mup\nu_\mu$ decay is modelled with the recent form factor predictions of Ref.\cite{McLean}. The main background originates from $\Bs$ semimuonic decays to excitations of the $\Dsm$ meson, with the dominant $D_s^{*-}\to \Dsm\gamma$ decay represented by a specific shape, and higher excitations $D_s^{**-} = [D^{*-}_{s0}(2317),D^-_{s1}(2460),D^-_{s1}(2536)]\to \Dsm X$ modelled by a combined shape. Other sources of background are the decays of the form $B\to \Dsm D X$ and the semitauonic decay $\Bs\to \Dsm\taup(\to\mup\nu_\mu\bar{\nu}_\tau)\nu_\tau$. Due to similarity of their shapes, the $\Bs\to D_s^{**-}\mup\nu_\mu$ channels are grouped with $B_s\to \Dsm D X$ decays, while \mbox{$\Bs\to \Dsm\taup(\to\mup\nu_\mu\bar{\nu}_\tau)\nu_\tau$} is combined with $B_{u,d}\to \Dsm D X$ decays.

\begin{figure}
	\begin{center}
	\includegraphics[width=0.49\textwidth]{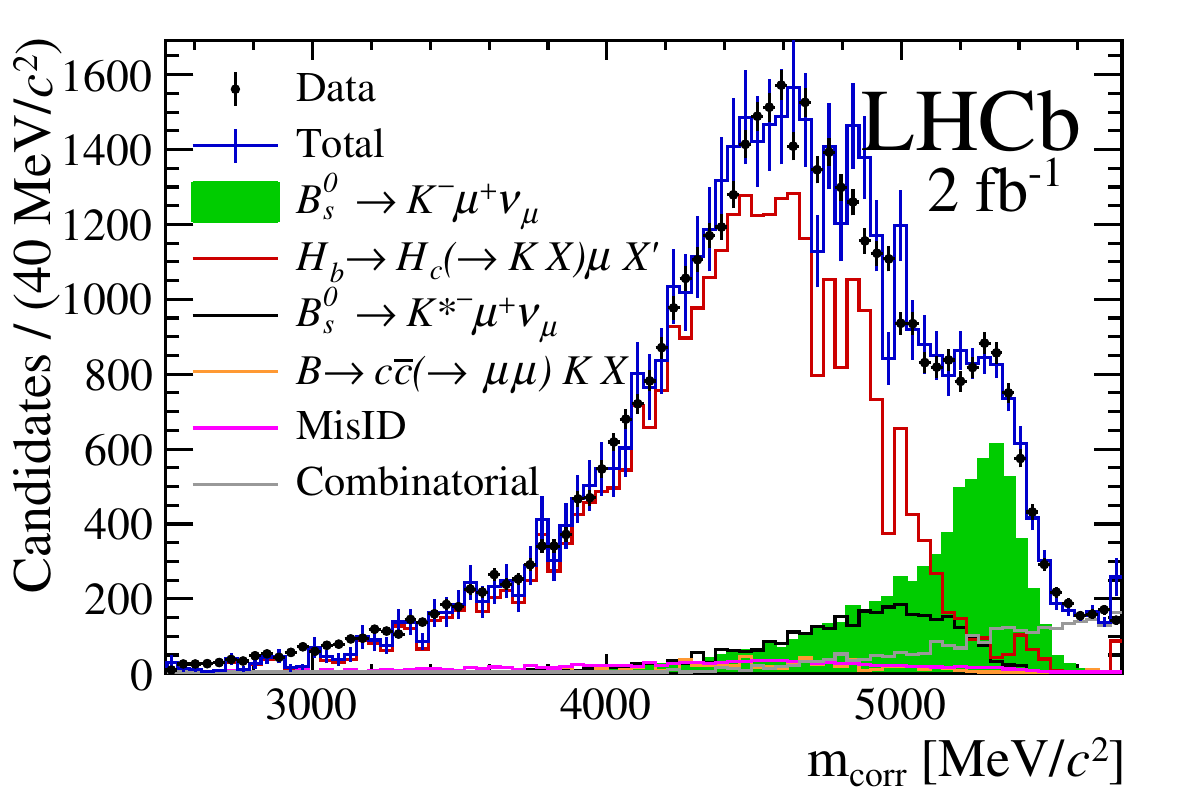}
	\includegraphics[width=0.49\textwidth]{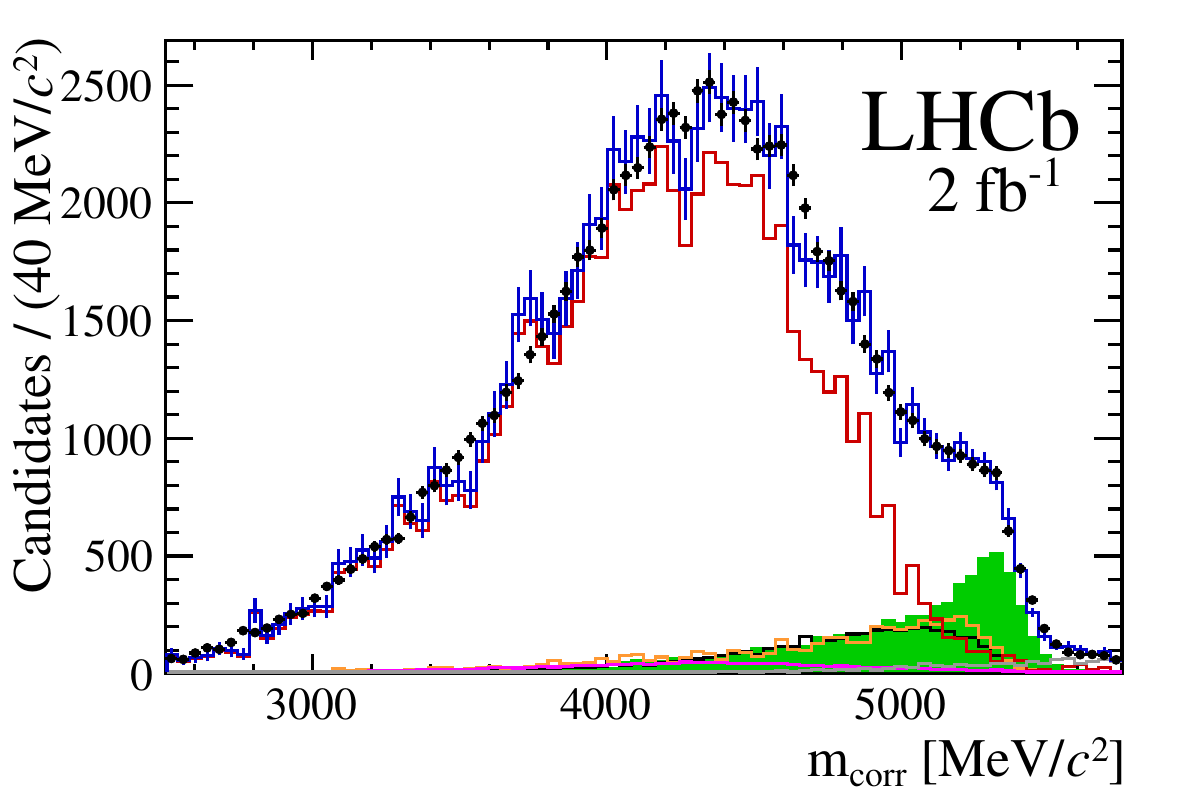}
        \end{center}
         \begin{center}
	\includegraphics[width=0.49\textwidth]{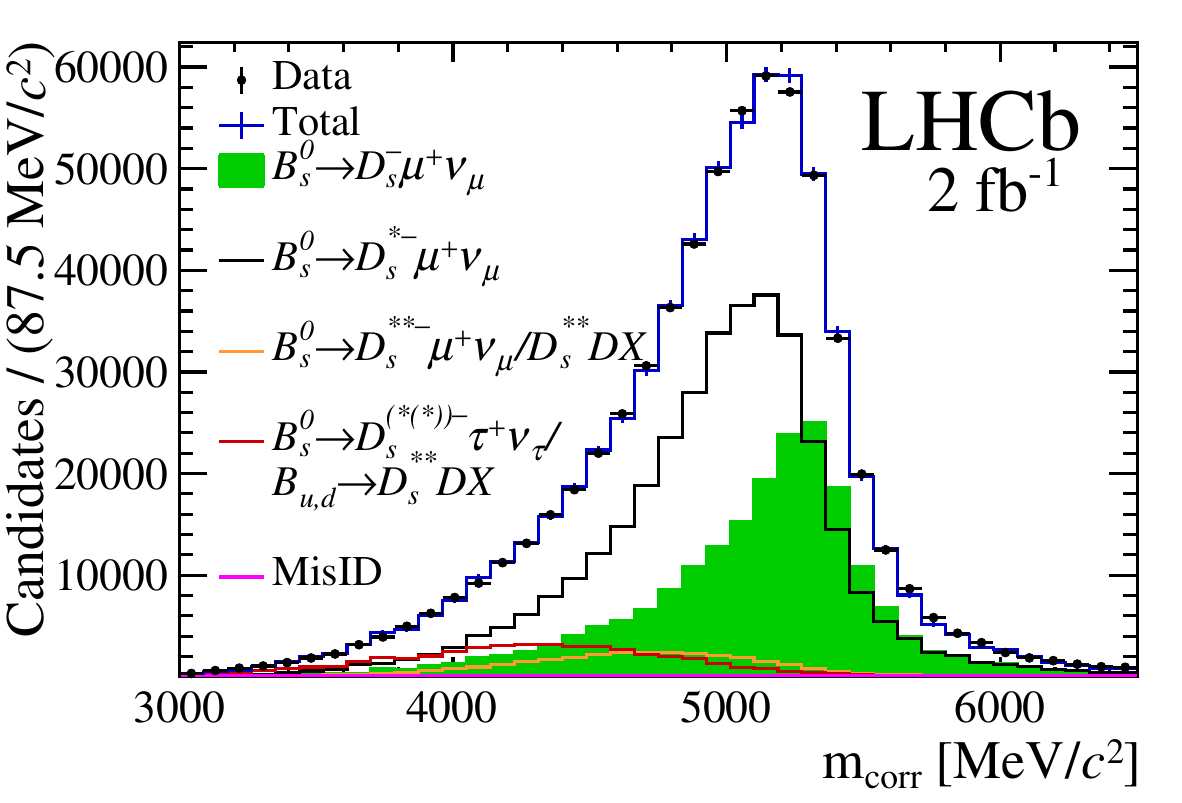}
         \end{center}
	\caption{Distribution of $m_{\rm corr}$ for (top) the signal $\Bs\to \Km\mup\nu_\mu$, with (left) $\qsq<7\gevgevcccc$ and (right) $\qsq>7\gevgevcccc$, and (bottom) the normalization $\Bs\to \Dsm\mup\nu_\mu$ channel.  The points represent data, while the resulting fit components are shown as histograms.}
	\label{fig:FITMCORR}
\end{figure}

The corrected mass distributions of the signal and normalization candidates are shown in Fig.~\ref{fig:FITMCORR}, with the binned maximum-likelihood fit projections overlaid. The $\Bs\to \Km\mup\nu_\mu$ yields for $\qsq<7\gevgevcccc$ and  $\qsq>7\gevgevcccc$ regions are found to be $N_K=6922\pm285$ and $6399\pm370$, respectively, while the $\Bs\to \Dsm\mup\nu_\mu$ yield is $N_{D_s}=201450\pm 5200$. The uncertainties include both the effect of the limited data set and the finite size of the samples used to derive the fit templates. Unfolding the two effects in quadrature shows that they have similar sizes.

This is the first observation of the decay $\Bs\to \Km\mup\nu_\mu$. The ratio of branching fractions is inferred as
\begin{equation}
R_{\mathrm{BF}}\equiv\frac{\mathcal{B}(\Bs\to \Km \mup\nu_\mu)}{\mathcal{B}(\Bs\to \Dsm \mup\nu_\mu)} = \frac{N_K}{N_{D_s}}\frac{\epsilon_{D_s}}{\epsilon_K}\times\mathcal{B}(\Dsm\to \Kp\Km\pim),
\end{equation}
with $\mbox{$\mathcal{B}(\Dsm\to \Kp\Km\pim)=(5.39\pm0.15)\%$}$ \cite{PDG2020} and gives
\begin{align*}
R_{\mathrm{BF}}(\mathrm{low})&=(1.66\pm 0.08\stat\pm0.07\syst\pm 0.05~(D_s))\times 10^{-3},\\
R_{\mathrm{BF}}(\mathrm{high})&=(3.25 \pm 0.21\stat^{\,+\,0.16}_{\,-\,0.17}\syst\pm 0.09~(D_s))\times 10^{-3},\\ 
R_{\mathrm{BF}}(\mathrm{all})&=(4.89\pm 0.21\stat^{\,+\,0.20}_{\,-\,0.21}\syst\pm 0.14~(D_s))\times 10^{-3}, 
\end{align*}
where the uncertainties are statistical, systematic and due to the $\Dsm\to \Kp\Km\pim$ branching fraction. \mbox{Table \ref{Tab:SystObs}} summarizes the systematic uncertainties. It includes uncertainties on the calibration and correction of the track reconstruction, trigger, particle identification, selection variables, migration of events between \qsq regions, efficiencies and the fit template distributions. The largest systematic uncertainty originates from the fit templates and is evaluated by varying the shape of the fit components according to alternative models and also by modifying within its uncertainty the mixture of exclusive decays representing some of the background contributions. In particular, the signal shape is varied using various form factor models~\cite{Bouchard:2014ypa,Flynn:2015mha,Bazavov:2019aom,Khodjamirian:2017fxg}. A similar procedure is applied to the normalization channel. The tracking uncertainty comprises the limited precision on tracking efficiency corrections obtained from control samples in data, and the uncertainty on modelling the hadronic interactions with the detector material. The uncertainty on the \qsq migration is related to the limited accuracy of the evaluation of the cross-feed between low and high \qsq regions in simulation.
\begin{table}
\centering
	\caption{Relative systematic uncertainties on the ratio $\mathcal{B}(\Bs \to \Km\mup\nu_\mu)/\mathcal{B}(\Bs \to \Dsm\mup\nu_\mu)$, in percent.}\label{Tab:SystObs}
	\begin{tabular}{l|ccc}
        \hline
	Uncertainty                      & All \qsq & low $\qsq$            & high $\qsq$\\
	    \midrule
	Tracking                               &2.0                 & 2.0 & 2.0 \\
        Trigger                                &1.4                & 1.2 & 1.6 \\
	Particle identification                &1.0                & 1.0 & 1.0\\
	$\sigma(m_\mathrm{corr})$                &  0.5                 &0.5 & 0.5           \\
	Isolation                              &0.2                &0.2 & 0.2  \\
	Charged BDT                            & 0.6                  &0.6   & 0.6\\
	Neutral BDT                          & 1.1                  &1.1   & 1.1\\
	\qsq migration                         &    --               &  2.0 & 2.0     \\
	Efficiency                &1.2                &1.6 & 1.6     \\
        Fit template                           &${}^{+2.3}_{-2.9}$     &${}^{+1.8}_{-2.4}$ &${}^{+3.0}_{-3.4}$\\
        \midrule                                                    
        Total                                 &${}^{+4.0}_{-4.3}$     &${}^{+4.3}_{-4.5}$ &${}^{+5.0}_{-5.3}$\\
        \midrule\hline
	\end{tabular}
\end{table}

To determine the branching fraction $\mathcal{B}(\Bs \to \Km\mup\nu_\mu)$ and the ratio $|V_{ub}|/|V_{cb}|$, the predicted integrals of the form factors $\mbox{$\mathrm{FF}_Y = |V_{x b}|^{-2}\int \frac{d\Gamma(\Bs\to Y\mup\nu_\mu)}{dq^2} dq^2$}$ \mbox{($Y=\Km, \Dsm$;} \mbox{$x=u,c$)} are required. The absolute branching fraction is calculated as \mbox{$\mathcal{B}(\Bs\to \Km\mup\nu_\mu)=\tau_{B_s}\times |V_{cb}|^2\times \mathrm{FF}_{D_s}\times R_{\mathrm{BF}}$}. The inputs are the exclusive value of $|V_{cb}|=(39.5\pm0.9)\times 10^{-3}$ \cite{PDG2020}, the \Bs meson lifetime $\tau_{B_s}=1.515\pm0.004\ps$ \cite{PDG2020} and the form factor integral $\mathrm{FF}_{D_s}=9.15\pm0.37\invps$ based on a recent LQCD computation \cite{McLean}. This leads to
\begin{equation}
\nonumber \mathcal{B}(\Bs\to \Km\mup\nu_\mu)=(1.06\pm0.05~(\mathrm{stat})\pm0.04~(\mathrm{syst})\pm0.06~(\mathrm{ext})\pm0.04~(\mathrm{FF}))\times 10^{-4},
\end{equation} 
where the uncertainties are statistical, systematic, from the external inputs ($\Dsm$ branching fraction, $\Bs$ lifetime and $|V_{cb}|$) and the $\Bs\to \Dsm$ form factor integral, respectively. Combining the systematic uncertainties, the branching fraction is \mbox{$\mathcal{B}(\Bs\to \Km\mup\nu_\mu)=(1.06\pm0.05\stat\pm0.08\syst)\times 10^{-4}$}.

\begin{figure}
	\centering
	\includegraphics[width=0.6\textwidth]{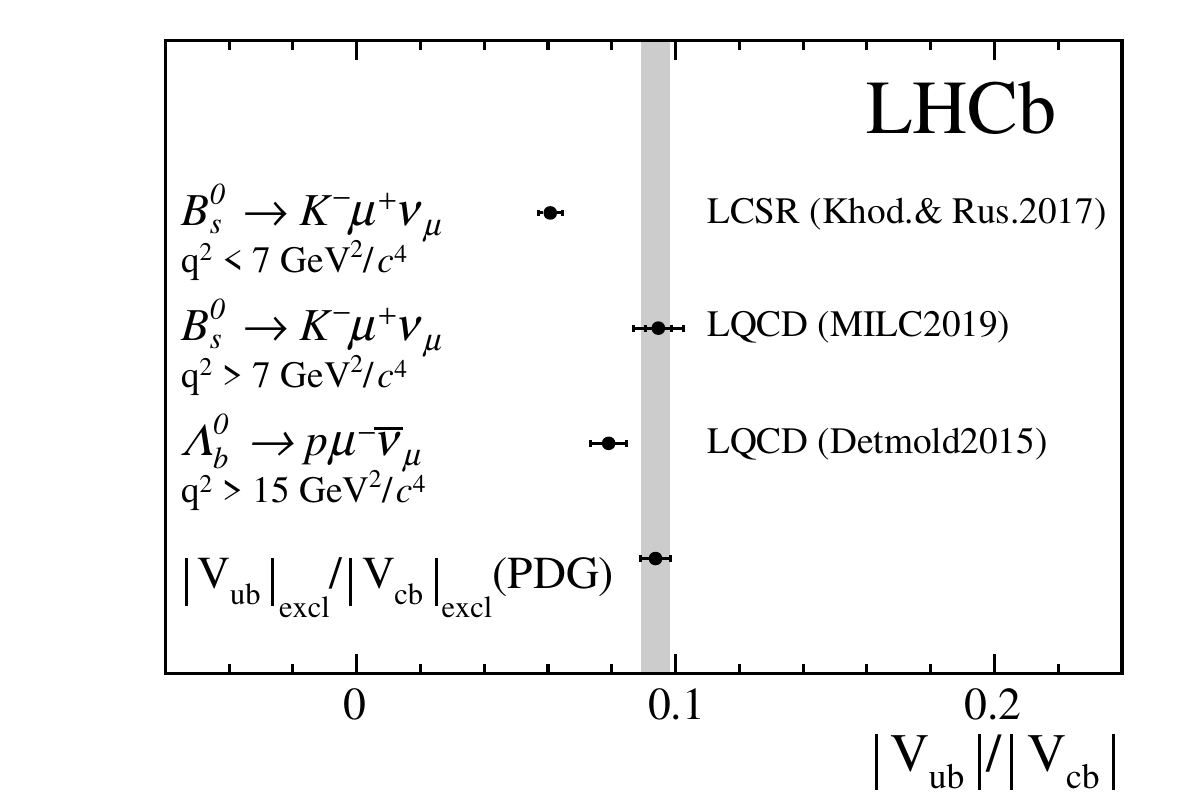}
	\caption{Measurements of $|V_{ub}|/|V_{cb}|$ in this Letter and in Ref.\cite{LHCb-PAPER-2015-013}, and ratio inferred from the PDG \cite{PDG2020} averages of exclusive $|V_{ub}|$ and $|V_{cb}|$ measurements, where the $\Lb\to p\mun\bar{\nu}_\mu$ result is not included. The form factor calculation used in each case is mentioned \cite{Khodjamirian:2017fxg,Bazavov:2019aom,Detmold:2015aaa}.}
	\label{fig:VubVcb_plot}
\end{figure}

 The ratio of CKM elements $|V_{ub}|/|V_{cb}|$ is obtained through the relation \mbox{$R_{\mathrm{BF}}= |V_{ub}|^2/|V_{cb}|^2 \times \mathrm{FF}_{K}/\mathrm{FF}_{D_s}$}. For the $\mathrm{FF}_K$ value, a recent LQCD prediction is used for the high \qsq range, $\mbox{$\mathrm{FF}_{K}(\qsq>7\gevgevcccc) = 3.32 \pm 0.46  \invps$}$ \cite{Bazavov:2019aom}, while a LCSR calculation \cite{Khodjamirian:2017fxg} is used for the low \qsq range, $\mbox{$\mathrm{FF}_{K}(\qsq<7\gevgevcccc) = 4.14 \pm 0.38 \invps$}$, due to the lower accuracy of LQCD calculations in this region.  
The obtained values are
\begin{align*}
{|V_{ub}|}/{|V_{cb}|}(\mathrm{low})&=0.0607\pm0.0015\stat\pm0.0013\syst\pm0.0008~(D_s)\pm0.0030~(\mathrm{FF}) ,\\
{|V_{ub}|}/{|V_{cb}|}(\mathrm{high})&=0.0946\pm0.0030\stat^{+~0.0024}_{-~0.0025}\syst\pm0.0013~(D_s)\pm0.0068~(\mathrm{FF}) , 
\end{align*}
where the latter two uncertainties are from the \Dsm branching fraction and the form factor integrals.
The discrepancy between the values of $|V_{ub}|/|V_{cb}|$ for the low and high \qsq ranges is related to the difference in the theoretical calculations of the form factors. To illustrate this, the LQCD calculation in Ref.~\cite{Bazavov:2019aom} gives $ \mathrm{FF}_{K} = 0.94 \pm 0.48  \invps$ at low \qsq, which can be compared to the chosen LCSR value, $4.14 \pm 0.38 \invps$ \cite{Khodjamirian:2017fxg}. Figure~\ref{fig:VubVcb_plot} depicts the $|V_{ub}|/|V_{cb}|$ measurements of this Letter, $\mbox{${|V_{ub}|}/{|V_{cb}|}(\mathrm{low})=0.061\pm0.004$}$ and $\mbox{${|V_{ub}|}/{|V_{cb}|}(\mathrm{high})=0.095\pm0.008$}$, with the uncertainties combined.  The ${|V_{ub}|}/{|V_{cb}|}$ measurement obtained with the $\Lb$ baryon decays \cite{LHCb-PAPER-2015-013}, for which a form factor model based on a LQCD calculation \cite{Detmold:2015aaa} was used, is also shown. 

In conclusion, the decay $\Bs\to \Km\mup\nu_\mu$ is observed for the first time. The branching fraction ratios in the two \qsq regions reported in this Letter represent the first experimental ingredient to the form factor calculations of the $\Bs\to \Km\mup\nu_\mu$ decay. Moreover, the ${|V_{ub}|}/{|V_{cb}|}$ results will improve both the averages of the exclusive measurements in the  $(|V_{cb}|,|V_{ub}|)$ plane and the precision on the least known side of the CKM unitarity triangle.

\section*{Acknowledgements}
%
%
\noindent We express our gratitude to our colleagues in the CERN
accelerator departments for the excellent performance of the LHC. We
thank the technical and administrative staff at the LHCb
institutes.
We acknowledge support from CERN and from the national agencies:
CAPES, CNPq, FAPERJ and FINEP (Brazil); 
MOST and NSFC (China); 
CNRS/IN2P3 (France); 
BMBF, DFG and MPG (Germany); 
INFN (Italy); 
NWO (Netherlands); 
MNiSW and NCN (Poland); 
MEN/IFA (Romania); 
MSHE (Russia); 
MICINN (Spain); 
SNSF and SER (Switzerland); 
NASU (Ukraine); 
STFC (United Kingdom); 
DOE NP and NSF (USA).
We acknowledge the computing resources that are provided by CERN, IN2P3
(France), KIT and DESY (Germany), INFN (Italy), SURF (Netherlands),
PIC (Spain), GridPP (United Kingdom), RRCKI and Yandex
LLC (Russia), CSCS (Switzerland), IFIN-HH (Romania), CBPF (Brazil),
PL-GRID (Poland) and OSC (USA).
We are indebted to the communities behind the multiple open-source
software packages on which we depend.
Individual groups or members have received support from
AvH Foundation (Germany);
EPLANET, Marie Sk\l{}odowska-Curie Actions and ERC (European Union);
A*MIDEX, ANR, Labex P2IO and OCEVU, and R\'{e}gion Auvergne-Rh\^{o}ne-Alpes (France);
Key Research Program of Frontier Sciences of CAS, CAS PIFI, CAS CCEPP, 
Fundamental Research Funds for Central Universities, 
and Sci. \& Tech. Program of Guangzhou (China);
RFBR, RSF and Yandex LLC (Russia);
GVA, XuntaGal and GENCAT (Spain);
the Royal Society
and the Leverhulme Trust (United Kingdom).




\addcontentsline{toc}{section}{References}
\bibliographystyle{LHCb}
\bibliography{main,LHCb-PAPER,LHCb-CONF,LHCb-DP,LHCb-TDR}

\newpage
\centerline
{\large\bf LHCb collaboration}
\begin
{flushleft}
\small
R.~Aaij$^{32}$,
C.~Abell{\'a}n~Beteta$^{50}$,
T.~Ackernley$^{60}$,
B.~Adeva$^{46}$,
M.~Adinolfi$^{54}$,
H.~Afsharnia$^{9}$,
C.A.~Aidala$^{85}$,
S.~Aiola$^{26}$,
Z.~Ajaltouni$^{9}$,
S.~Akar$^{65}$
% ...

{\footnotesize \it

% ...
$ ^{9}$Universit{\'e} Clermont Auvergne, CNRS/IN2P3, LPC, Clermont-Ferrand, France\\
% ...
$ ^{26}$INFN Sezione di Milano, Milano, Italy\\
% ...
$ ^{32}$Nikhef National Institute for Subatomic Physics, Amsterdam, Netherlands\\
% ...
$ ^{46}$Instituto Galego de F{\'\i}sica de Altas Enerx{\'\i}as (IGFAE), Universidade de Santiago de Compostela, Santiago de Compostela, Spain\\
% ...
$ ^{50}$Physik-Institut, Universit{\"a}t Z{\"u}rich, Z{\"u}rich, Switzerland\\
% ...
$ ^{54}$H.H. Wills Physics Laboratory, University of Bristol, Bristol, United Kingdom\\
% ...
$ ^{60}$Oliver Lodge Laboratory, University of Liverpool, Liverpool, United Kingdom\\
% ...
$ ^{65}$University of Cincinnati, Cincinnati, OH, United States\\
% ...
$ ^{68}$Syracuse University, Syracuse, NY, United States\\
% ...
$ ^{85}$University of Michigan, Ann Arbor, United States, associated to $^{68}$\\
\bigskip
% ...
\medskip
}
\end{flushleft}

\end{document}